%% file: main.tex
\documentclass[conference]{IEEEtran}
\usepackage{cite}
\usepackage{amsmath,amssymb,amsfonts}
\usepackage{algorithmic}
\usepackage{graphicx}
\usepackage{textcomp}
\usepackage{xcolor}
\usepackage[hyphens]{url}

\usepackage{mathptmx}
\usepackage[utf8]{inputenc}
\usepackage[T1]{fontenc}

\usepackage{fancyhdr}
\usepackage[normalem]{ulem}
\usepackage{xcolor}
\usepackage{tabulary}
\usepackage{paralist}
\usepackage[bookmarks=true,breaklinks=true,letterpaper=true,colorlinks,linkcolor=black,citecolor=blue,urlcolor=black]{hyperref}
\usepackage{subcaption}

\graphicspath{ {./graphs/}{./figures/} }

\newcommand{\new}[1]{{\color{blue}{#1}}}
\renewcommand{\new}[1]{#1}

\newcommand{\GS}{GETX\textit{spy}}

\def\smallerspacecaption{\vspace{-2mm}}

\newcommand{\iscasubmissionnumber}{916}

\title{Coherence Attacks and Countermeasures in Interposer-Based Systems}
\author{\normalsize{ISCA 2022 Submission
    \textbf{\#\iscasubmissionnumber} -- Confidential Draft -- Do NOT Distribute!!}}

\begin{document}

  \author{  \IEEEauthorblockN{Gino Chacon}
            \IEEEauthorblockA{\textit{ginochacon@tamu.edu}\\
                                \textit{Texas A\&M University} }
  \and
            \IEEEauthorblockN{Tapojyoti Mandal}
            \IEEEauthorblockA{\textit{tapojyoti.mandal@tamu.edu}\\
                                \textit{Texas A\&M University}}\\

            \IEEEauthorblockN{Paul V. Gratz}
            \IEEEauthorblockA{\textit{pgratz@gratz1.com} \\
                                \textit{Texas A\&M University}}
 \and
            \IEEEauthorblockN{Johann Knechtel}
            \IEEEauthorblockA{\textit{johann@nyu.edu} \\
                                \textit{New York University Abu Dhabi}}\\

	 \IEEEauthorblockN{Vassos Soteriou}
            \IEEEauthorblockA{\textit{NA} \\
                                \textit{Cyprus University of Technology}}
 \and
            \IEEEauthorblockN{Ozgur Sinanoglu}
            \IEEEauthorblockA{\textit{ozgursin@nyu.edu} \\
                                \textit{New York University Abu Dhabi}}
 }
%
%
%
%
%

\maketitle
\thispagestyle{plain}
\pagestyle{plain}

\maketitle
\input{abstract}
\input{intro}
\input{background}
\input{arch_overview}
\input{example_attack}
\input{threat_model}
\input{design}

\input{evaluation}

\input{conc}



\bibliographystyle{IEEEtranS}
\bibliography{ref}

\end{document}

%% file: abstract.tex
\begin{abstract}
  Industry is moving towards large-scale systems where processor
  cores, memories, accelerators, etc.\ are bundled via 2.5D
  integration.  These various components are fabricated
  separately as chiplets and then integrated using an interconnect
  carrier, a so-called interposer.  \new{This new design style
    provides benefits in terms of yield as well as economies of
    scale, as chiplets may come from various third-party vendors, and
    be integrated into one sophisticated system.  The benefits of this approach,
    however, come at the cost of new challenges for 
    the system's security and integrity when many third-party
    component chiplets, some from not fully trusted vendors, are
    integrated.}

  %

  Here, we explore these challenges, but also promises, for modern
  interposer-based systems of cache-coherent, multi-core chiplets.
  First, we \new{introduce a new, coherence-based attack, \GS, wherein
    a single compromised chiplet can expose a high-bandwidth
    side/covert-channel in an ostensibly secure system.  We further
    show that prior art is insufficient to stop this new
    attack.}  Second, we propose using an active interposer as
  generic, secure-by-construction platform that forms a physical root
  of trust for modern 2.5D systems.  Our scheme has limited overhead,
  restricted to the active interposer, allowing the chiplets and the
  coherence system to remain untouched.  We show that our scheme
  prevents a wide range of attacks\new{, including but not limited to
    our \GS\ attack,} with little overhead on system performance,
  $\sim$4\%.  This overhead reduces as workloads increase, ensuring
  scalability of the scheme.
\end{abstract}

%% file: intro.tex
\section{Introduction}
\label{Introduction}

A recent trend in computing systems is the adoption of 
hardware organization based on chiplets and
interposers~\cite{vivet20,kim19,naffziger20,naffziger21}. Instead of
implementing a monolithic system-on-chip (SoC), this approach
disaggregates the functional components across multiple smaller chips,
i.e., \textit{chiplets}, which are designed and manufactured
separately.  These chiplets serve as hard intellectual property (IP) modules, possibly sourced from a
variety of vendors, and consolidated on an integration and
interconnects carrier, i.e., the
interposer~\cite{vivet20,kim19,naffziger20,naffziger21,matsuo00,takaya13}.
This approach is also known as 2.5D integration.

The adoption of chiplet and interposer integration raises design reuse
to the level of the physical system, optimizing yields and
streamlining time to market, resulting in significant cost benefits.
Such 2.5D integration is already adopted by industry in products such
as the AMD Epyc processors~\cite{naffziger20,naffziger21} or the Intel
Embedded Multi-Die Interconnect Bridge
technology~\cite{7545486}. \new{Recent industry talks have herald this
  design style as the next iteration of Moore's
  law~\cite{intel_ceo_interview}.}

\begin{figure}
\centering
  \begin{subfigure}{.48\columnwidth}
    \centering
    \includegraphics[width=\columnwidth]{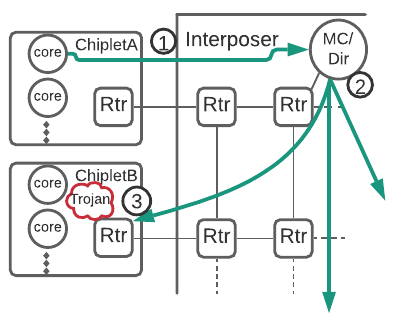}
    \caption{\textbf{Passive Reading:} Trojan passively observes write
      traffic for other chiplets.  \textbf{(1)} Misses from Chiplet A
      cause \textbf{(2)} broadcast invalidations to all chiplets;
      \textbf{(3)} Trojan snoops invalidation addresses. Our \GS\
      attack is an example of this threat.}
    \label{fig:passive_reading_attack}
  \end{subfigure}
  \hfill
  \begin{subfigure}{.48\columnwidth}
    \centering
    \includegraphics[width=\columnwidth]{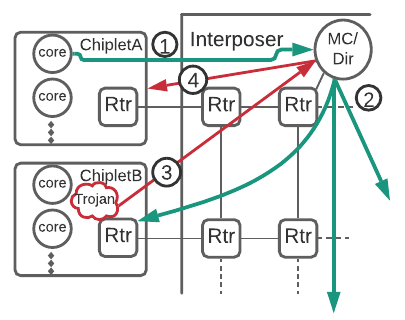}
    \caption{\textbf{Masquerading:} Trojan acts as another
      core. \textbf{(1)} Miss causes GETX to directory; \textbf{(2)}
      broadcast invalidations to each chiplet; \textbf{(3)} Trojan
      blocks local observation, forges reply with different core ID;
      \textbf{(4)} requesting core proceeds, leaving local caches
      incoherent.}
    \label{fig:masquerade_attack}
  \end{subfigure}
  \begin{subfigure}{.48\columnwidth}
      \centering
      \includegraphics[width=\columnwidth]{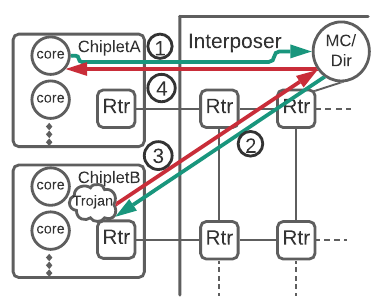}
      \caption{\textbf{Modifying:} Trojan forges message to achieve
        incoherent state. \textbf{(1)} Chiplet A sends GETS to
        directory; \textbf{(2)} directory forwards request to Trojan's
        core which has line in `E' state.  Trojan blocks GETS and
        \textbf{(3)} replies with GETX to requestor, \textbf{(4)}
        invalidating Chiplet A's cache entry, leaving the attacker in
        control of another cache's contents.}
      \label{fig:modifying_attack}
  \end{subfigure}  
  \hfill
  \begin{subfigure}{.48\columnwidth}
    \centering
      \includegraphics[width=\columnwidth]{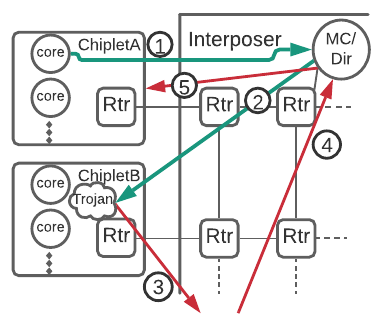}
      \caption{\textbf{Diverting:} Trojan diverts invalidation
        requests. \textbf{(1)} Chiplet A sends GETX to the directory;
        \textbf{(2)} directory broadcasts invalidations. \textbf{(3)}
        Trojan blocks message and diverts a request to another core,
        \textbf{(4)} which responds with a negative-acknowledge or
        acknowledgment resulting in \textbf{(5)} the directory
        allowing original requestor to continue.}
      \label{fig:divert_attack}
  \end{subfigure}
  \caption{Examples of coherence-oriented Trojan attacks in
    interposer-based systems. In each case, Chiplet A is the victim of
    a Trojan attack from Chiplet B.}
\label{fig:coherence_attacks}
\smallerspacecaption
\smallerspacecaption
\end{figure}

\subsection{Security Challenges}

Interposer-based systems are vulnerable to not only traditional
attacks, but also a range of dedicated, new attacks. For example,
vulnerabilities may be introduced through the various third-party
chiplets, e.g., via untrusted fabrication~\cite{5604161} of chiplets,
malicious or simply buggy third-party IPs~\cite{rajendran16} within
chiplets, or collusion of multiple malicious actors across
chiplets. If not addressed properly, the vulnerability of a single
chiplet may undermine the entire system's security.

Coherence is an essential mechanism which ensures all components
maintain a consistent view of memory, not only for interposer-based
systems, but interconnected SoCs in general.
%
\new{The predictability and prevalence of the coherence system makes
  it an attractive target, yet only a few related attacks have been
  proposed~\cite{Yao_2018,Trippel_2018,7926975}, e.g., a Trojan interacting
  with the coherence system can allow attackers to gain, or
  deny~\cite{7926975}, control of the memory system or grant
  privilege-escalated access.  Integrating defenses into the
  coherence system is a difficult task that requires extensive
  verification and design effort.  Defenses that (naively) interact
  with the coherence system may cause functional bugs and deadlocks.}

\new{While prior work in secure network-on-chip (NoC) fabrics consider
  untrusted IP modules,
  they do not address the full scope of a coherence-oriented attack.
  These defenses are generally limited to the detection of attacks,
  limited to a single class of attack~\cite{Boraten_GLSVLSI_2016,
    Raparti_2019}, fail to prevent attacks against coherence-system
  interactions~\cite{Florin_2008,Prodromou_2012, Saeed_2014}, require
  additional complex hardware~\cite{Prodromou_2012, Atul_2019,
    LemayG14}, or require packet authentication through
  error-correction codes~\cite{Boraten_DATE_2016} or key
  exchanges~\cite{Charles_2020,7951731,Gebotys_2003} which increases
  network bandwidth pressure. In contrast, our scheme removes the
  burden of securing the NoC itself and allows for defensive
  strategies that target securing the coherence-level communications.}



\new{Figure~\ref{fig:coherence_attacks} outlines a subset of attacks
  that a hardware Trojan can mount against a hybrid
  broadcast/directory coherence protocol, e.g.,
  MOESI Hammer~\cite{Conway_2010}. While selected and simple examples, these 
  represent severe threats for interposer-based systems. We note that 1)~these
    attacks can be applied to any broadcast protocol and 2)~to
    apply them in directory protocols, the Trojan would 
    first need to fake a GETS to register the lines locally in the
    directory, ensuring that invalidations would be sent to the
    Trojan; doing so is valid and practical.}

  The threat landscape is further demonstrated through a new 
  covert-channel attack, \GS.  This channel
  is established by a \textit{spy} process running on one chiplet that
  may otherwise remain uncompromised. The \textit{spy} process
  exfiltrates data from its chiplet via specific, but legal,
  write-ownership coherence requests (GETX). These requests are
  observed by a Trojan snooping the coherence messages from the cache
  controller in a second, compromised chiplet's core.  Although we
  formulate this attack as a covert-channel, the underlying mechanism
  can be trivially refactored into a side-channel attack, leaking the
  addresses of cachelines written by the victim chiplet.  We
  demonstrate that this covert-channel allows for covert transmission
  at rates of up to \textit{4.22Mbps}.

\GS\ applies to any 2.5D system that enforces coherence, but is unique
to such hardware orchestration because integrating chiplets from
different vendors increases the risks for Trojan exposure and multiple
actors maliciously colluding across chiplets.  \new{Prior works do not
  address such attacks enabled by \textit{legal} cache coherence interactions.} Note
that the goal of this work is to protect not only against this
particular covert-channel attack, but against all other system-level
threats arising from untrusted chiplets integrated into a 2.5D
systems.




Establishing some ``root of trust'' is critical to ensure the
security and integrity of data in modern systems containing various
third-party IP components and software applications interacting on the
system.  Commercial solutions such as ARM's
TrustZone~\cite{TrustZone_2009} and Intel's SGX~\cite{ SGX_2016}, as
well as academic proposals~\cite{7951731, maene18, zhang19,
  lebedev19}, typically rely on dedicated microarchitectural support
and other measures, e.g., memory encryption.  These approaches often
incur high overheads and are prone to dedicated
attacks~\cite{sec17-lee-jaehyuk,qiu19}, while being susceptible to
hardware Trojans throughout the outsourced supply
chains~\cite{BT18,mehta20}, a fact often overlooked in prior art.
%

\subsection{Security Promise of Interposers, Our Contributions}

We leverage the notion of interposer-based system design to establish
a secure-by-construction root of trust in \new{modern multi-core,
  multi-chiplet systems}. Importantly, unlike prior art for secure
system design, \textit{we do not assume/require trusted manufacturing
  of the whole system, only of the interposer, to provide system-level
  security promises.}

We introduce a security-centric interconnect fabric within an active
interposer, which monitors and controls all system-level communication
with low performance overheads.  Our design does not interfere with
the system's underlying coherence protocol, but rather prevents
sensitive information from being divulged to, or manipulated by,
untrusted chiplets.

The contributions of our work are as follows:
\begin{enumerate}

\item We examine how chiplets of an interposer-based system can be
  attacked a)~directly via unprivileged memory references,
  b)~indirectly via attacks conducted at the NoC level, namely
  unauthorized access, snooping, spoofing, modifying, or diverting of
  messages, and c)~indirectly via covert-channels.  For the latter, we
  demonstrate a simple but effective attack that allows
  a hardware Trojan to receive messages from a spy process operating
  in another chiplet. \new{\textit{No prior work has identified this
      kind of attack and existing security mechanisms cannot mitigate it.}}

\item To protect against these threats around \new{coherence-oriented}
  system-level communication, \textit{we propose an active interposer
    as the physical backbone for a secure-by-construction root of
    trust in multi-chiplet systems.}

\item We introduce a novel microarchitecture to secure communication
  passing from untrusted chiplets onto the interposer (and thus into
  the system) based upon per-packet validation at the interposer
  ingress links.  Our design does not modify the underlying coherence
  system, but rather prevents it from exposing sensitive information.
  \textit{The key objective of our proposal is to realize a secure
    large-scale system out of untrusted chiplets.}

\item We implement our proposed technique and examine the implications
  of our security features.  We characterize the performance impact
  as a low, $\sim$4\% overhead. Further, we show the overhead
  decreases as workloads scale.

\end{enumerate}

We note that developing a secure operating system (OS) for our system
is outside the scope of this work.  Prior work in secure OS
and virtualization systems~\cite{Witchel_2005, Chen_2008,
  Costan_2016,Koning_2017} or systems described in
Sec.~\ref{sec:rot} may be extended accordingly.

%% file: background.tex
\section{Background, Motivation, and Contributions}
\label{BackgroundMotivation}
Here, we review key concepts of interposer technology, hardware security,
and cache coherence protocols.
%
We also motivate the contributions of our work
considering the security challenges and promises for the respective 
state-of-the-art.

\subsection{Interposer Technology}
\label{sec:interposerbasic}

Interposer technology, also known as 2.5D integration, is the process
of manufacturing two or more chips, or chiplets, separately and
subsequently integrating and interconnecting them using a carrier made
of silicon or other
materials~\cite{vivet20,kim19,naffziger20,naffziger21,matsuo00,takaya13}.
Compared to traditional, monolithic SoC designs, 2.5D integration
drastically reduces time to market. A system designer can procure IP
as commodity chiplets and directly integrate them at the physical
system level, with effort only required for designing the interposer.
2.5D integration is beneficial in general, as it allows for design and
manufacturing process optimization, increasing yield for chiplets.
Although future  2.5D designs will be more heterogeneous, current
state-of-the-art systems are largely homogeneous, cache-coherent, multi-core
chiplet designs~\cite{vivet20,coudrain19,kim19,naffziger20,naffziger21}.

%

Active interposers contain active devices (e.g., NoC
routers, voltage regulators, sensors, etc.), while passive interposers
act solely as an integration carrier and wiring medium.  Although
passive interposers are cheap to manufacture, their physical design
can be quite challenging~\cite{jerger15,kim19}.  In contrast to active
interposers with buffering of interconnects, passive interposer wires are of
considerable length, incurring significant power and delay overheads.
An active interposer with an embedded NoC fabric serves well for
large-scale chiplet integration and system communication. The chiplet
interconnect fabric is encapsulated away from the interposer NoC
beyond the edge router on the interposer to which it is attached. Such
heterogeneous fabric allows for cross-optimization of topologies
across chiplets and interposer, opening up considerable opportunities
for system
design~\cite{jerger15,vivet20,coudrain19,activeNoC_2018isca,coskun20,Bharadwaj20}.
Further, active interposers improve
testability~\cite{takaya13,hellings15,vivet20} and thereby help to
manage the yield of the final system.

Active interposers are typically manufactured in relatively older
nodes~\cite{vivet20,Nabeel_2020}. Therefore, it is realistic that a
trusted facility is available for manufacturing of such active
interposers.  \textit{Here we propose an active interposer-based root of trust
with security features embedded within its NoC routers.}

\subsection{Hardware Security}

\subsubsection{IC Manufacturing}

Industry has widely adopted a work mode where IC design and
verification is carried out by a design house and partners, but
fabrication and testing is outsourced to off-shore facilities
typically providing access to advanced technologies. While this
practice reduces the cost of production and streamlines the time to
market~\cite{7827668}, it raises concerns regarding the
trustworthiness of the outsourced fabrication facilities, which may
seek to insert security vulnerabilities in general or hardware Trojans
in particular~\cite{5604161}.

The threat vector posed by untrusted fabrication facilities implies
the ICs they manufacture are untrustworthy.  This causes a security
challenge for modern systems in multiple ways.  First, any hardware
security feature embedded in such outsourced IC may no longer offer
the desired protection, presenting a profound challenge.  Second, a
modern system may be composed of
chiplets 
with various levels of trustworthiness. Any malicious chiplet behavior
may compromise the whole system due to its interconnected nature.

The interposer technology can help to avoid such complications.
This is because an interposer can be fabricated
separately in a trusted facility and may also embed security features.
Accordingly, \textit{an interposer designed to constitute a hardware-enforced
root of trust can be built upon to ensure the overall system's
trustworthiness, as we show in this work.}

\subsubsection{Hardware Trojans}
Hardware-centric attacks such as the malicious insertion or
modifications of circuitry, also known as hardware Trojans, can lead
to catastrophic security failures within a system. For example,
Bidmeshki et al.~\cite{7753274} provide an attack scenario wherein a
hardware Trojan renders the cryptography subsystem vulnerable, Khan
et al.~\cite{9061138} demonstrate Trojans that can leak data from
cache memory of processors, and Kim et al.~\cite{7926975} introduce
Trojans which inject malicious coherence messages to create a
denial-of-service attack.

Our work is orthogonal to and compatible with prior art on Trojan
detection and mitigation, e.g., \cite{wu21, trippel20, guo19_QIFV}.
We do not seek to prevent Trojans, rather to prevent their attacks
from affecting the system-level security.  Specifically, we seek to
prevent any hardware-centric attacks that are executed through the
memory and coherence system.  This notion of system-level security is
enforced by a clear physical separation of untrusted commodity
chiplets and security features residing in the trusted
interposer. \textit{Prior art on Trojan detection and mitigation cannot offer
such secure-by-construction organization.}

\subsubsection{Secure Interconnect Fabrics}
Prior art for NoC security assumes that malicious activities arise
from connected components or the network fabric itself.  Fiorin et
al.~\cite{4492766} propose security features for policy-based message
checking against untrusted components.  Selected works focus on
securing the system through encryption/decryption of packets/messages
exchanged through NoC fabrics~\cite{Gebotys_2003, Evain_2005}.  Kinsy
et al.~\cite{7951731} propose organizing secure and non-secure
software/hardware entities as tenants and configure the NoC routers to
securely exchange messages.  While enabling a secure NoC fabric, the
amount of key exchanges required to isolate nodes/tenants incurs high
latencies and is not easily scaled.


Nabeel et al.~\cite{Nabeel_2020} propose an interposer-based
architecture where security modules monitor the interconnect fabric at
the level of bus addressing, to block transactions that violate memory
access policies.  While their design represents a relevant first work
toward secure 2.5D integration, it has several limitations.  First,
the authors consider an overly simplistic architecture, ignoring the
fact that state-of-the-art 2.5D designs are fully memory-mapped and
cache-coherent.  We find addressing the coherence model is critical
to providing system-level security.  Second, the authors did overlook
new security challenges arising for interposer designs.  Critically,
their design would fail to hinder the \GS\ cover/side-channel attack
we study in this work, as \GS\ does not violate memory access policies/permissions.
%
%

\new{For most prior art, networks are not secure-by-construction,
  hence high-overhead solutions are required such as key-based
  security~\cite{Charles_2020, Florin_2008, Raparti_2019}, model
  checking~\cite{Prodromou_2012, Boraten_2016}, or additional
  structures to verify traffic
  patterns~\cite{Prodromou_2012,Atul_2019,LemayG14}.  While
  packet-checking schemes similar to our design have been proposed in
  the past, e.g., \cite{Saeed_2014}, the underlying defense mechanisms
  often address only a single attack
  vector~\cite{Boraten_GLSVLSI_2016, Raparti_2019} and/or fail to
  address the coherence system's exploitable
  nature~\cite{Florin_2008,Prodromou_2012, Saeed_2014}. While these
  works check the message's memory operation, they do not
  differentiate between specific coherence message types and the ways
  that coherence messages can be exploited beyond simple read or write
  traffic.  Even more concerning, most prior art assumes, often
  implicitly, trusted manufacturing of the whole system.  Such an
  assumption is challenged by outsourced supply chains.
  These concerns are only exacerbated for 2.5
  integration using chiplets from various vendors.}

By contrast, our work does not make such overarching assumptions. \textit{We
enforce system-level security for untrusted commodity chiplets by
integrating them on an interposer-based root of trust, the only
component requiring trusted fabrication, thereby providing a
secure-by-construction NoC.} Without the need to secure the integrity
of the NoC, a more simplified approach may be taken to ensure the
security of the overall system, resulting in lower overheads.

\subsubsection{Hardware Support for Root of Trust}
\label{sec:rot}
Intel's SGX provides an extension to create trusted execution
environments (TEEs), called enclaves~\cite{SGX_2016,
  SGX_EXPLAINED_2016}. Enclaves prevent unprivileged access to secure
data during security-sensitive execution.  Specifically, SGX maps
protected memory pages to reserved memory regions in which the pages
are encrypted by a hardware encryption module.  However, recent work
shows vulnerabilities in SGX, due to programming errors and untrusted
software~\cite{Khandaker_2020, Park_2020}, as well as due to
speculative execution~\cite{Bulck_2018, Schwarz_2019, Chen_2019,
  Mohammadian_2018}.  ARM's hardware-enforced TEE isolates secure
execution from untrusted software~\cite{TrustZone_2009}. AMD's TEE
leverages a normal OS running in tandem with a
secure OS. The latter has access to the full range of a device's
peripherals and memory, whereas the normal OS only has access to a
subset of peripherals and memory regions, to prevent unauthorized
access of sensitive resources. However, recent work shows TEEs are
prone to vulnerabilities due to architectural, implementation, and
hardware issues~\cite{Cerdeira_2020}.

In short, these schemes incur high overheads, are prone to dedicated
attacks, and are susceptible to Trojans in general. In contrast, \textit{our
approach has little impact on system performance and its key
components are secure-by-construction.}

\subsection{Cache Coherence}

Coherence protocols ensure updates to cached copies of data are
visible to all cores in modern multi-core designs~\cite{coudrain19,
  vivet20, naffziger21}.  Broadly speaking, coherence schemes can be
categorized as broadcast (or snooping) protocols~\cite{Bilir_1999,
  Sinharoy_2005,Agarwal_2009} and directory
protocols~\cite{Archibald_1984, Zebchuk_2009, Laudon_1997}.  Broadcast
protocols, while simple to implement, suffer from high traffic due to
the amount of messages multi-core systems require to maintain
coherence. Directory protocols allow for fine-grained state tracking
and unicast messages, making them highly scalable, but difficult to
implement and have higher access latencies.

A coherence protocol is generally oblivious to the software and may
permit malicious accesses that leak sensitive
information~\cite{Yao_2018,Trippel_2018}.  Existing countermeasures
address conflict-based and transient-execution side-channel attacks,
but do not consider threats from maliciously manipulated/malformed
coherence message packets~\cite{Yan_2017, KirianskyLebedev_2018,
  Yan_2019, YanWen_2019}.  Given that the coherence protocol acts only
based on rules for how memory is updated across multiple parties,
attackers may exploit the coherence protocol's low-level behavior. We
demonstrate one such attack in Sec.~\ref{sec:getx_spy}.  It is
important to note that coherence is a hardware-managed,
micro-architectural feature which is neither influenced by, nor
exposed to, the software executing on the system, rendering
software-based defenses ineffective.


Our solution does not require modifying the coherence 
protocol (which would impose extensive verification efforts and can
result in complex, adversarial side-effects for the system behavior).
Rather, \textit{we carefully ensure messages' integrity and prevent untrustworthy
chiplets from exploiting the coherence protocol and system-level memory 
management.}


%% file: arch_overview.tex
\section{System Architecture Overview}
\label{Arch}

\begin{figure}[tb]
\centering
\includegraphics[width=\columnwidth]{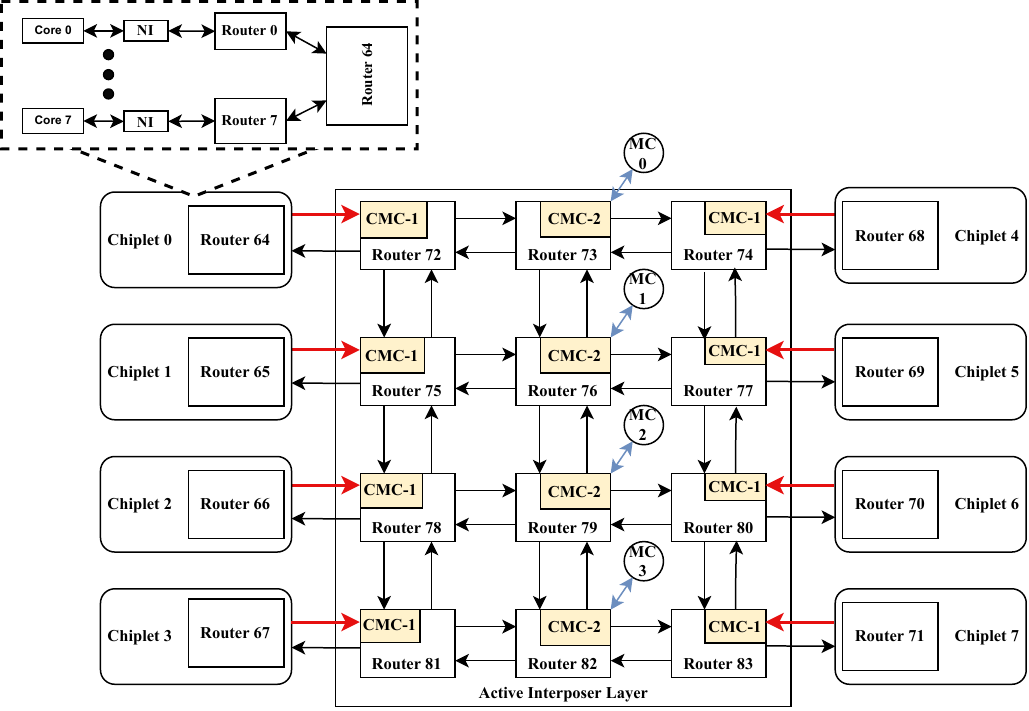}
\caption{Baseline system architecture.  Routers 64--71 lie within 
  chiplets, connecting them to the interposer NoC.  Routers 0--63
  connect the CPU cores within their respective chiplet's NoC
  (see also zoom-in).  The proposed CMCs, marked in yellow,
  are placed along the ports/links connected to chiplets (CMC-1, red
  arrows) and memory controllers (CMC-2, blue arrows).  
}
\label{fig:System_Architecture}
\smallerspacecaption
\smallerspacecaption
\end{figure}

Figure~\ref{fig:System_Architecture} outlines the secure,
interposer-based, multi-chiplet and multi-core system proposed in this
work.  The baseline system is loosely based on the architecture of the
Rocket-64 design proposed by Kim et al.~\cite{kim19}.  In addition to
the overview in this section, more details are provided in
Sec.~\ref{sec:design}.

\subsection{Chiplet and Interconnects Architecture}
In this system, we employ eight chiplets, each containing eight CPU
cores, for 64 cores in total, similar to recent AMD
processors~\cite{naffziger20,naffziger21}.%
\footnote{Our proposed architecture places no restrictions on the
  contents of the chiplets; be they cores, accelerators, GPUs, etc.,
  as long as they are cache-coherent and obey shared-memory semantics.
  Here we focus on homogeneous chiplets, representing the state-of-the-art design style.
 Also, this scenario makes simulation in gem5~\cite{lowepower2020gem5} more practical.}
  Each core has an L1 instruction and data cache, and a
unified L2 cache; all cache levels are private to each core.  The
cache controllers generate coherence messages which the network
interface (NI) in each chiplet converts to network packets prior to
injection into the interposer NoC (via interface routers). Chiplets
are interconnected to each other and to four memory controllers (MC)
via an NoC of 2D mesh topology residing in the active interposer. The
interface routers, depicted along the east and west edges of the
system, serve as ingress links for the chiplets into the interposer
NoC.

Many other architectures are practical for interposer-based
systems~\cite{jerger15,vivet20,coudrain19,activeNoC_2018isca,coskun20,Bharadwaj20}.
Assuming a cache-coherent shared memory and an active interposer, our
proposed system can be readily ported to such.  Furthermore, the
security principles leveraged in our work---verification and
policy-checking of memory-system messages, as outlined below and
further detailed in Sec.~\ref{sec:design}---are extendable to various
physical fabrics and communication protocols in homogeneous and
heterogeneous systems. For example, interfaces such as PCIe are
typically memory-mapped; checking of memory-system messages can
prevent unauthorized access by any malicious chiplets. Although some
heterogeneous systems may not fully enforce coherence, memory-system
messaging is still used to communicate between processing elements and
I/O, etc.

\subsection{Principles and Features for System-Level Security}



We propose the interposer as root of trust for 
integration of untrustworthy chiplets into a secure system, namely by
enforcing policy checking of all system-level communication.  The key
attributes to enable such a secure system are: (1)~the
interposer is manufactured separately from the untrusted chiplets, in
a trusted facility; and (2)~the interposer serves as integration
and communication backbone between chiplets.

Any system-level
communication across chiplets must pass through the interposer.
Accordingly, all memory traffic must traverse the interposer NoC as
network packets and are checked before entering the network. If a CPU
core wants to read/write data from/to memory, a corresponding
coherence message, embedded in a packet, must traverse the interposer
NoC. Similarly, if a core wants to communicate with another core in
another chiplet, such direct messages must also traverse the
interposer NoC.  Importantly, \textit{all direct communication
  messages are limited to legal coherence messages}, as is typical in
most multi-processor systems.  Thus, we embed related security
features exclusively within the interposer NoC such that all messages
must inevitably traverse through, and be checked by, the trusted
active interposer.

We add \emph{Coherence Message Checkers (CMCs)} to the physical
ingress links 
to validate all coherence messages coming from
the chiplets into the active interposer. We also add CMCs to the
physical links connecting with the MCs; the related details are
discussed in Sec.~\ref{sec:CMC_types}. \textit{Since CMCs are implemented
exclusively within the trusted active interposer, their hardware is
trustworthy and free from Trojans by construction.}


%
%

\subsection{Cache Coherence Protocol}
\label{sec:cc}

We focus on the \textit{MOESI Hammer} cache coherence
protocol~\cite{Conway_2010} as basis for our implementation, which is
used in many AMD systems as scalable protocol for multi-core
systems. Our approach, however, is easily extendable to other
coherence schemes as well.

MOESI Hammer is a hybrid protocol; it encapsulates the
scalability of directory-protocols without high implementation
complexity while achieving the low-latency of broadcast protocols
without overly increasing broadcasted coherence message traffic.  To
that end, MOESI Hammer maintains a sparse directory between multiple
home nodes to track cache lines' states and owners.  Coherence
requests access a cache line's home-node directory and DRAM in
parallel to reduce the cost of a directory miss, cancelling the DRAM
response if a directory entry is found. Traffic is reduced by only
broadcasting to all cores for specific state transitions.

We note that coherence broadcast protocols increase exposure to Trojan
attacks, which might snoop broadcast messages to memory regions
otherwise inaccessible.  Our security approach addresses this threat
directly and in a novel way.  Next, we showcase one such concrete
attack, and then we follow-up with our threat model, which covers
other common threats for system-level security of modern interconnected designs.

%% file: example_attack.tex
\section{The \GS\ Attack}
\label{sec:getx_spy}


Here we introduce and demonstrate a new attack leveraging the
vulnerability of the coherence mechanism to hardware Trojans in
untrusted chiplets.
  Specifically, our attack a)~can transmit any
data from one chiplet, via a regular user process acting as
\textit{spy} that generates tailored write-ownership coherence
messages (GETX), and b)~employs a hardware Trojan in a compromised
chiplet that passively reads those GETX requests.  We call the attack
\GS\ as it relies on GETX requests generated by the spy.  \new{No
  prior security scheme we are aware of is capable of preventing this
  kind of attack. That is because \GS\ observes the addresses of legal
  invalidation messages; it does not violate the system's coherence
  protocol, allowing it to evade the defense mechanisms of prior
  work.}

While the attack demonstration is specific to MOESI
  Hammer, the working principle can be easily applied to various broadcast or
  directory protocols in interposer-based systems.
Also, while this attack
focuses on the threat of a compromised coherence system in
interposer-based designs, note that our proposed scheme also prevents further, more
generic attack vectors outlined in Sec.~\ref{Threat Model} and studied
in Sec.~\ref{Evaluation}.



\subsection{Working Principle}

MOESI Hammer (Sec.~\ref{sec:cc}) uses a coarse-grained directory
distributed between multiple memory controllers (MCs). Each core has
its own local directory to maintain coherence.
When an MC directory receives a GETX request without an
existing entry, a broadcast message is sent to all cores. This
expected interaction can be used to create a simple covert-channel
between a spy process and a hardware Trojan placed at the cache
controller directory in one of a chiplet's cores to receive
information via broadcasted GETX messages.

While our attack exploits the coherence state of specific addresses,
similar to~\cite{Yao_2018}, it differs in a few important aspects.
First and most importantly, \GS\ does not require the spy and Trojan
to operate within the same virtual address space.  Second, our
covert-channel does not rely on a Trojan process to query the targeted
addresses.  Third, our attack is not reliant on timing memory
accesses. Finally, our Trojan is simply a malicious observer of memory
requests, representing a realistic and concerning scenario that is
hard to mitigate.

\begin{figure}[tb]
  \begin{center}
    \includegraphics[width=.98\columnwidth]{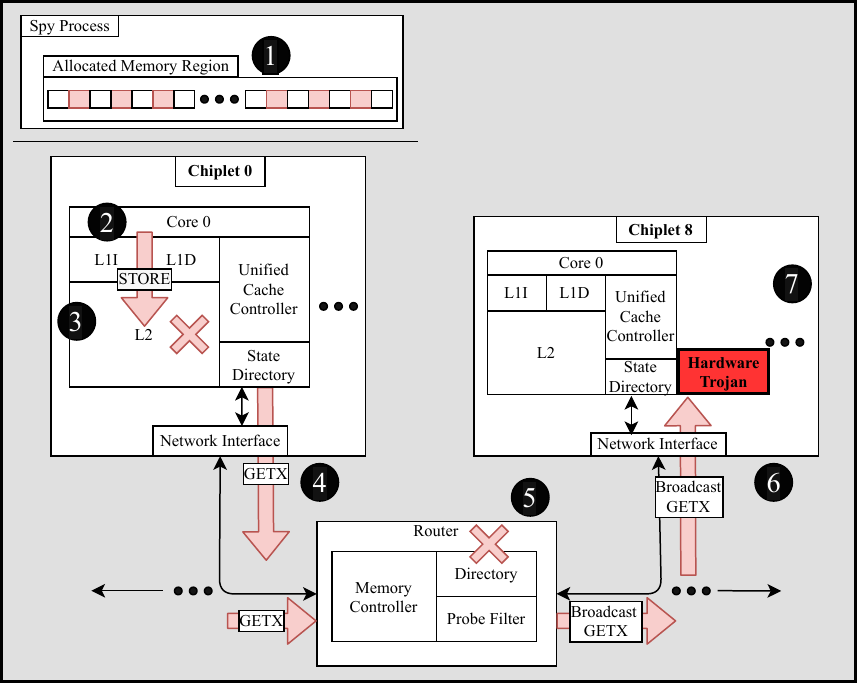}
    \caption{The \GS\ attack, executed as spy process in Chiplet 0's core 0,
    sending covert-channel messages to the hardware Trojan located in Chiplet 8's core 0. }
    \label{fig:Attack_Diagram}
  \end{center}
\smallerspacecaption
\smallerspacecaption
\smallerspacecaption
\end{figure}


Figure~\ref{fig:Attack_Diagram} illustrates the attack
orchestration. \textbf{(1)}~The spy process allocates a large memory
region to continually cause remote requests without pausing to flush
the L2.  \textbf{(2)}~The spy writes to targeted sets, causing misses
in the L2.  \textbf{(3)}~Each miss generates a new GETX request.
\textbf{(4)}~The GETX is sent to the MC to check for a directory entry
or ``hit'' in the probe filter.  \textbf{(5)}~The GETX misses in the
MC, resulting in a broadcast GETX to invalidate any shared copies of
the data present in other cores.  \textbf{(6)}~The chiplet containing
the hardware Trojan receives the broadcasted GETX, which buffers the
request. Using the L2 set index bits, the Trojan checks if a
synchronization message has been received.  \textbf{(7)}~After
synchronization, the Trojan observes GETX requests from the spy
process to receive covert messages.

Critically, the chiplet holding the Trojan (Chiplet 8 here) does
not need shared access to the spy process's virtual address range, as
the coherence protocol mandates GETX requests be broadcast to all
cores, \emph{regardless of physical page ownership}.  Also, the attack
does not require the memory region to be primed or the caches to be
flushed before a new transmission.

\new{\GS\ can be trivially reworked into a
  side-channel attack.  In such case, the \GS\ Trojan would
  passively watch for GETX-induced, invalidation broadcasts, to
  spy on the write address patterns of processes in other chiplets.
  Here again, the chiplet containing the Trojan need not have any
  access to the virtual address space or physical pages of the
  processes being spied upon.}


\subsection{\GS\ Case Study}

We implement \GS\ on the system described in
Sec.~\ref{Arch}, for the unsecured baseline version versus the
proposed secure version.  The evaluation setting is described in
Sec.~\ref{Evaluation}.  The system has an 8-way associative L2 cache
and 4-way associative MC directory. This means targeting 32 addresses,
16 for each set representing `1' or `0' to transmit, allows for
continuous flushing of the L2 target sets (covert transmission).

\begin{figure}[tb]
\hrule
\vspace{0.5em}
  \begin{subfigure}{\columnwidth}
    \centering
      \includegraphics[width=0.98\textwidth]{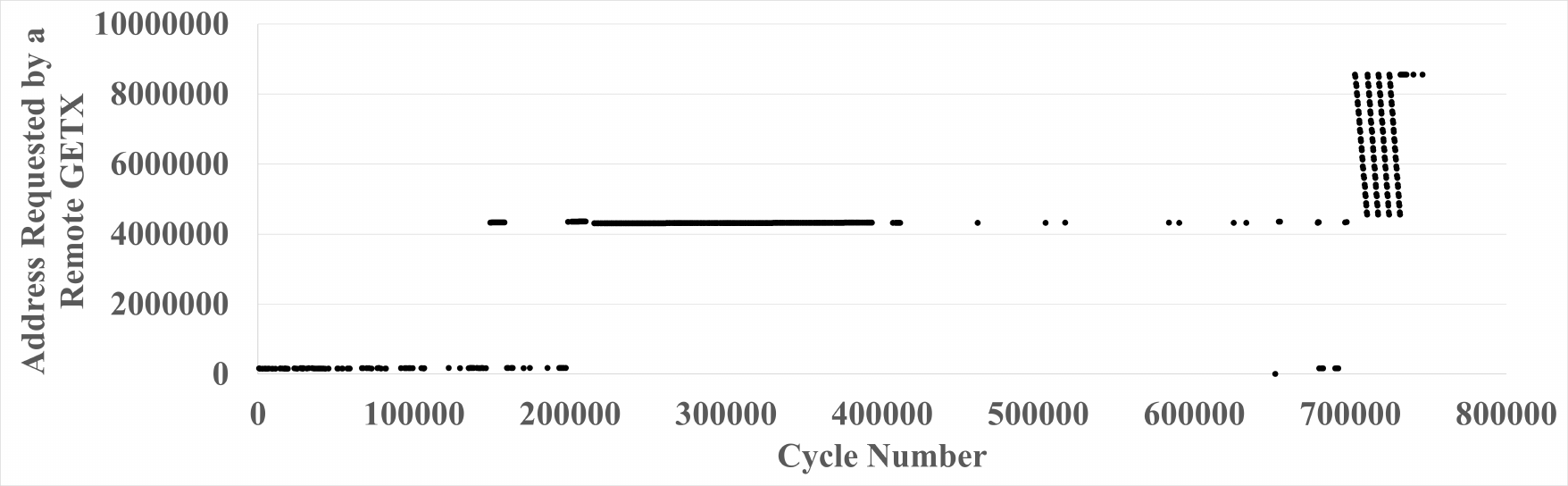}
      \caption{Addresses the hardware Trojan sees, as GETX requested
        from the spy process.  The attack occurs later in execution,
        when the spy targets specific addresses, to trigger misses in
        the L2 and the MC's directory.}
      \label{fig:getx_addresses}
  \end{subfigure}
\vspace{0.5em}
\hrule
\vspace{0.5em}
\begin{subfigure}{\columnwidth}
    \centering
    \includegraphics[width=0.98\textwidth]{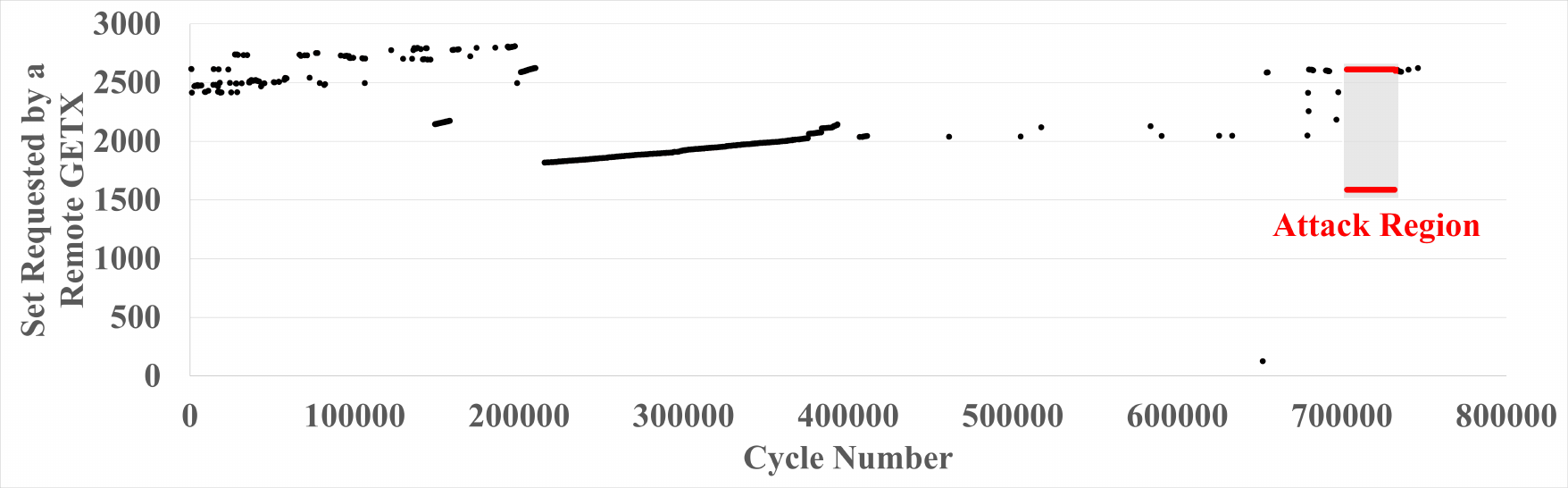}
    \caption{As the Trojan observes addresses requested, it awaits a
      synchronization message pattern, labelled as ``Attack Region.''}
    \label{fig:getx_sets}
\end{subfigure}
\vspace{0.5em}
\hrule
\vspace{0.5em}
\begin{subfigure}{\columnwidth}
    \centering
    \includegraphics[width=0.98\textwidth]{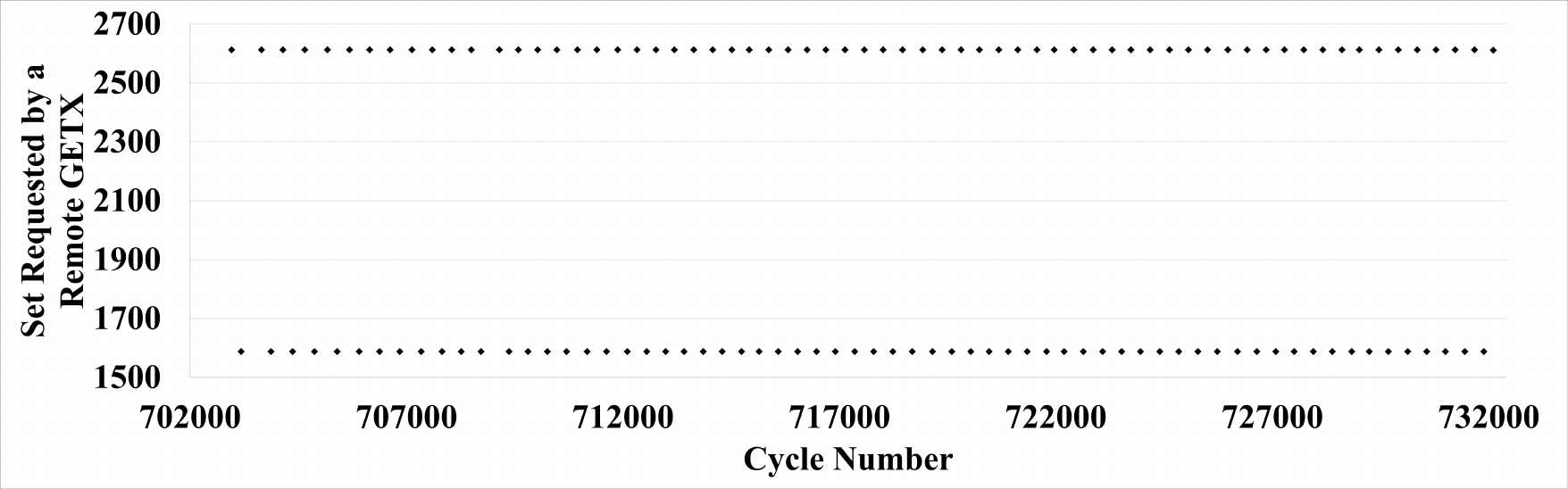}
    \caption{The attack region is zoomed-in here, showing the sets the
      Trojan considers as part of a synchronization message. The
      higher set represents `1' bits and the lower set represents `0'
      bits.}
    \label{fig:getx_attack_region}
\end{subfigure}
\vspace{0.5em}
\hrule
\vspace{0.5em}
\begin{subfigure}{\columnwidth}
    \centering
    \includegraphics[width=0.98\textwidth]{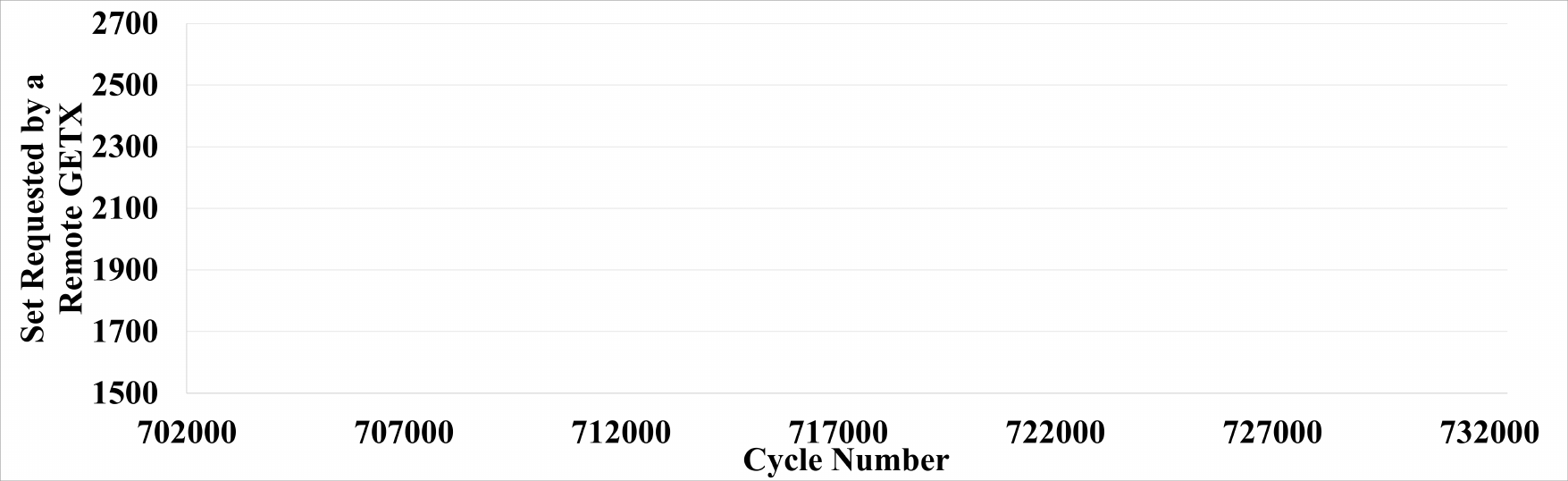}
    \caption{The same zoomed-in attack region seen by the Trojan while
      running on the proposed secure system. The Trojan can no longer
      observe any GETX messages sent by the spy process.}
    \label{fig:getx_attack_defended}
\end{subfigure}
\vspace{0.5em}
\hrule
\vspace{0.5em}
\caption{GETX requests, broken down into the address sets they
  reference and those the Trojan considers as part of the
  covert-channel transmission from the spy.}
\smallerspacecaption
\smallerspacecaption
\end{figure}

Figure~\ref{fig:getx_addresses} shows the addresses requested by the
spy process via GETX, as seen by the Trojan within a different core
and chiplet. The sets referenced by the addresses are shown in
Fig.~\ref{fig:getx_sets}. The spy process performs requests to
allocate memory which are viewed by the hardware Trojan as
inconsistent accesses and therefore considered irrelevant. The attack
region that is seen by the Trojan, Fig.~\ref{fig:getx_attack_region},
shows the spy later sending requests between two distinct sets to
represent a `1' or `0,' respectively.  The graph in
Fig.~\ref{fig:getx_attack_defended} shows the GETX requests seen by
the Trojan---namely none---when the attack is executed on our proposed
secure design.

Table~\ref{tab:Channel_Characteristics} shows the characteristics of
the \GS\ attack's covert-channel for the unsecured
baseline system. With a 4.22 Mbps bandwidth, this covert-channel
provides the basis for executing a range of data-leakage
attacks or other threats.
%


\begin{table}[tb]
  \small
  \centering
  \begin{tabular}{c|c}
    \hline
      Message Size &128 bits \\ \hline
      Cycles Taken to Transmit&28924 \\ \hline
      Clock Frequency&1GHz \\ \hline
      Total Time Taken to Transmit&28.92$\mu$s \\ \hline
      Megabits per Second&4.22 \\ \hline
      Percentage of NoC Bandwidth&0.013\%
  \end{tabular}
  \caption{\GS\ Covert-Channel Characteristics}
  \label{tab:Channel_Characteristics}
\smallerspacecaption
\smallerspacecaption
\end{table}

Our proposed scheme prevents the \GS\ Trojan from observing
  requests originating from the spy, thereby thwarting the
  covert-channel. Importantly, blocking messages to a chiplet directly at the
  chiplet's network interface within the interposer,
  based upon the memory permissions of that chiplet, is generic
  and applicable to further threats as well, as outlined next.

%% file: threat_model.tex
\section{Threat Model}
\label{Threat Model}



The focus of this work is a system wherein multiple chiplets have been
fabricated in various facilities and then connected together using 
interposer technology. The assumption is that the fabrication
as well as operational behavior of the
chiplets, either designed in-house or composed of third-party IPs,
cannot be trusted.
In other words, we assume that some Trojan(s) may exist in some chiplet(s).\footnote{%
Our work is orthogonal to and compatible with prior art on Trojan
detection and mitigation, e.g., \cite{wu21, trippel20, guo19_QIFV}.  We do
not seek to prevent Trojans, but to prevent their attacks from
affecting the system-level security.}
%
We also assume that attacks are targeted at memory-system traffic
which is the only type of traffic physically passing through the interposer.

Attacks on interconnected systems can be broadly categorized as outlined in
~\cite{7833075}.  Accordingly, our model considers the related four types of 
threat vectors (shown in Fig.~\ref{fig:coherence_attacks}):

\textit{Passive reading, aka snooping}: This threat occurs when a
malicious chiplet can read data not meant to. The \GS\
attack demonstrated in Sec.~\ref{sec:getx_spy} is an example of such a
threat in that the Trojan monitors broadcasted GETX requests to snoop a 
tailored message. 

\textit{Masquerading, aka spoofing}: This threat occurs when a
malicious chiplet disguises itself as another chiplet to gain access
to sensitive data or control of resources. Malicious chiplets can
modify the requester IDs and memory addresses embedded in cache
coherence messages, tricking directories or other unsuspecting cores
into divulging sensitive data.

\textit{Modifying}: Such threats
modify cache coherence messages.  For example, a chiplet may attempt
to disguise itself as having write
access to a memory region it has only read access to.

\textit{Diverting}: In shared-memory applications, a malicious chiplet
may divert data meant for one chiplet to another untrusted
chiplet, bypassing memory permissions. It may also divert cache
coherence messages, undermining the protocol.

In general, we aim to prevent hardware- and software-driven unauthorized access to
memory regions at the chiplet granularity, whether by software
privilege escalation, transient execution attacks, cache
side-channels, or any other means.


Our scheme provides protection on a chiplet granularity.  Attacks
across cores but within the same chiplet~\cite{10.1007/11605805_1,
  190938}, are out of scope of this work.  Similarly, out of scope are
attacks wherein code running on one core attempts to violate the
security of other processes running on that same core or on another
core in the same chiplet.  Further, attacks wherein one chiplet can
leverage memory transactions to its assigned memory region to modify
DRAM rows that are not assigned to it, e.g.,
Rowhammer~\cite{kim2014flipping}, are out of scope.
We note that prior art for protecting against such threats is orthogonal to our 
work and can be applied in addition.

%% file: design.tex
\section{System Design}
\label{sec:design}

Our proposed design prevents
attacks running on any given chiplet from violating the security of
the overall system, as we physically enforce protection against any
unauthorized access to shared-memory regions and conduct continuous
checking of the integrity and validity of cache coherence messages.
Next, we discuss the system design.

\subsection{Microarchitecture}
\label{Impl}

\subsubsection{CMC Overview}
With the proposed CMCs, we monitor and validate all incoming packets to
the interposer.
Figure~\ref{fig:CMC-1 placement} depicts the CMC embedded in a router
of the interposer NoC.  The CMC monitors messages traversing the
physical links prior to entering the virtual channel buffers within
the routers. Each CMC has two components described as follows:

\begin{figure}[tb]
  \begin{center}
    \includegraphics[width=\columnwidth]{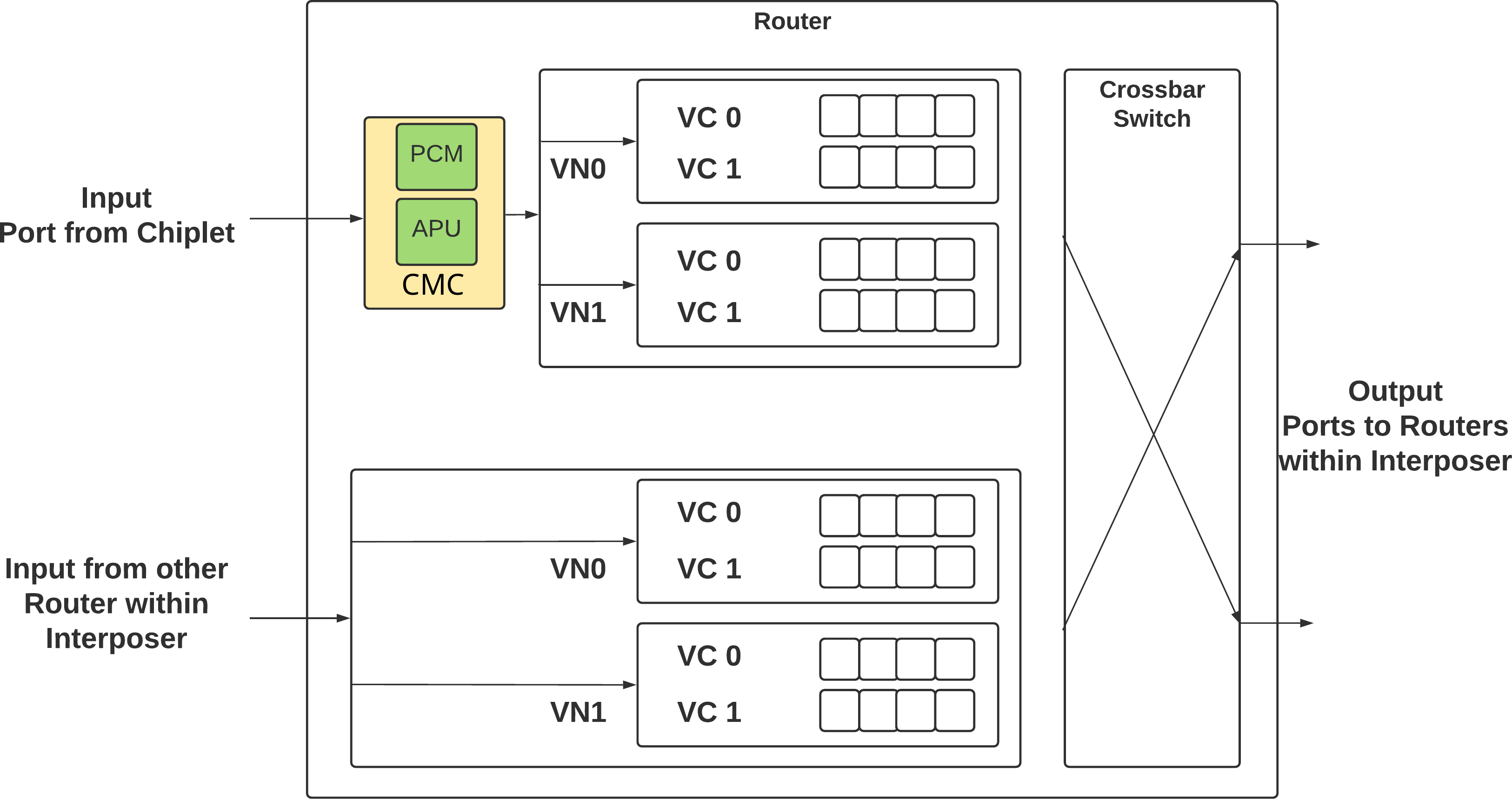}
    \caption{A CMC, embedded within an interface router of the interposer NoC,
      monitoring the incoming packets.}
    \label{fig:CMC-1 placement}
  \end{center}
\smallerspacecaption
\smallerspacecaption
\smallerspacecaption
\end{figure}

\textit{Packet Checker/Modifier (PCM)}: The PCM monitors and modifies
cache coherence messages as needed. Because the proposed system
  follows standard shared-memory semantics, all legal communication
  between cores, other IPs, I/O buses, and memory occurs through
  memory accesses which create cache coherence messages. Thus, the
PCM operates on coherence messages to check addresses and permissions;
modifying messages as needed.  More details are discussed in
Sec.~\ref{sec:Implementation}.

\textit{Address Protection Unit (APU) Table}: This is a direct-mapped,
SRAM-based look-up table with entries for each memory region and their
associated per-chiplet permissions.  As outlined in
Sec.~\ref{sec:smsa}, the main physical memory is partitioned into
multiple fixed-size regions. Each memory region has a corresponding
entry in the APU; hence, the number of entries within the APU table is
determined by the number of regions in the main memory.


\subsubsection{CMC Types and Placement}
\label{sec:CMC_types}

Recall Fig.~\ref{fig:System_Architecture}, depicting CMCs embedded in
the secure interposer-based system. The CMCs connected to the physical
links coming from chiplets are denoted as ``CMC-1''
and those connected to the physical links for MCs are denoted
``CMC-2.''  CMC-1 only monitors and verifies coherence messages
entering the interposer, whereas CMC-2 modifies certain coherence
messages at the directories (to counter passive-reading threats on
broadcast messages).  Router-to-router connections running exclusively
within the trusted interposer do not require CMC monitoring.

\textit{CMC-1}: Prevents the attached chiplet from injecting malicious
coherence messages into the system that violate the provisions of the
shared-memory organization, as outlined in Sec.~\ref{sec:smsa}.  The
PCM within CMC-1 monitors all traffic from the attached chiplet based
on the physical address the packet refers to. This physical address is
compared against the per-region permissions stored in the APU table
(described further below). If a message is of an allowed type to an
allowed memory region for the given chiplet (e.g., a GETX to a
read-only memory region it owns), the message may proceed into the
interposer NoC. Otherwise, if the packet is rejected, a dedicated
security signal, realized as a machine-check exception, is thrown and
system execution stops.\footnote{This is a secure and protocol-conform
  approach.  For the sake of system-level throughput, one may want to
  only isolate the chiplet(s) triggering a security violation. Doing
  so safely, however, is not trivial, as it would require significant
  modifications of the coherence protocol itself to prevent
  deadlocks. }

\textit{CMC-2}: Prevents the broadcast of coherence messages to
chiplets which are not permitted to access the related memory regions.
As described in Sec.~\ref{sec:cc}, MOESI Hammer does not maintain
per-core sharing information, hence certain message requests cause the
directory to broadcast the request to all cores.  The cores then
respond based on whether the cache block is shared by that core.  This
raises a concern of passive reading/snooping; recall the \GS\ attacks
in Sec.~\ref{sec:getx_spy}.

To prevent snooping, the PCM determines whether a given broadcast
message is directed towards a chiplet allowed to access the referred
memory region (based on the APU table). If the chiplet does not have
access, the broadcast message is converted into an appropriate
response message directed only to the original requester.  This is
legal within the coherence scheme of the system: if a chiplet is not
allowed to access a memory region, then its caches cannot contain
lines associated with that region.  This allows the CMC-2 to safely
divert broadcast messages from the directory and prevent snooping.


\subsubsection{APU Table}


The APU table is a lookup table containing entries describing the
access permissions for applications running within a respective
chiplet. Each entry corresponds to a pre-determined physical memory
region.  The access permissions are determined by a secure OS that is
running exclusively within the active interposer, independently of the
regular OS running on the chiplets. The permissions are programmed
into the APU tables during runtime, as outlined in
Sec.~\ref{sec:smsa}.

\begin{figure}[tb]
  \begin{center}
    \includegraphics[width=\columnwidth]{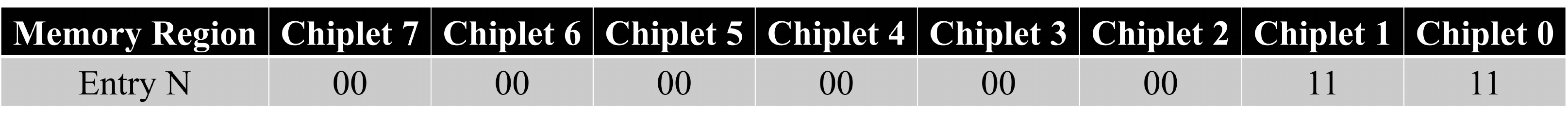}
    \caption{Exemplary entry of the APU table, covering some region of
      the physical memory.  The entry describes access permissions for
      each chiplet individually; here, the related region is
      read-write shared between Chiplets 0 and 1.  }
    \label{fig:APU_Table}
  \end{center}
\smallerspacecaption
\smallerspacecaption
\smallerspacecaption
\end{figure}
  
Figure~\ref{fig:APU_Table} details one entry in the APU table.  Each entry 
represents one memory region, with two bits allocated per chiplet to represent 
the access permissions of applications running in the chiplet: a chiplet may 
have no access permissions (`00'), read-only permissions (`01'), or read/
write permissions (`11'). The encoding `10' is unused.

When the PCM intercepts a packet, the upper bits of its physical
address are extracted and used to index into the APU table.  The
related entry is read and handed back to the PCM to compare the
request type, requester ID, and destination ID against the permission
levels in the APU table entry.

\subsection{OS Support and Shared-Memory Organization}
\label{sec:smsa}

Designing a secure OS to capitalize on our active interposer-based
root of trust is beyond the scope of this work.  However, prior work
in security-enabled OS environments~\cite{Witchel_2005, Chen_2008,
  Costan_2016,Koning_2017} and TEEs (Sec.~\ref{sec:rot}) may be
extended accordingly.  The interposer may include a trusted
co-processor to support a secure OS, secure boot-up and execution
environments~\cite{TrustZone_2009,Woodruff_2014,Hua_2017}.  In our
scheme, critical tasks like updating the APU table must be delegated
to such a secure environment, as the chiplets are physically unable to
access the proposed security features. That is, attacks on the APU
table and other components are prevented by construction.

In shared-memory systems, permissions are typically defined per
physical page by the OS during memory allocation.  Enforcing per-page
permissions in a CMC poses several challenges. Specifically,
page-level tracking requires a TLB-like structure to cache
translations~\cite{Witchel_2002}.  The required support for
maintaining the structure in coherence with the full system's page
table significantly increases hardware complexity and performance
overhead.  We argue that a page-level implementation at the interposer
is excessive in a system of relatively few and coarse-grained
chiplets.  Instead, we partition physical memory into coarse-grained
memory regions, similar to prior art~\cite{Lampson_1974, Witchel_2005,
  Woodruff_2014,Koning_2017}.

We aim for a ``sweet spot'' between too coarse-grained, where only
few memory regions are available and capacity is wasted to
fragmentation, versus too fine-grained, where the APU table could not
hold the excessive number of regions without incurring high access
latency or placing entries in a backstore.  We find that a total
number of memory regions between 4x--8x the number of chiplets is more
than sufficient for most use-cases, allowing for diverse private and
shared memory regions without too much fragmentation.
	  
Each memory region is designated as read- or write-able independently
to any given chiplet, with permissions updated as needed.  Data
private to a single chiplet is placed in a region (or set of regions)
only accessible by that chiplet. A page shared across multiple
chiplets is assigned to a memory region the given chiplets are allowed
access to.

Initial memory partitioning and permission setting occurs during the
initial soft page fault on a virtual page. Region allocations and
permissions are updated via an API call from the OS, e.g., similar to
Intel's SGX page allocation model~\cite{SGX_2016}.  After a page
fault, the OS requests a memory range for the process from the secure
OS operating on the interposer. The APU table for the chiplet that
requested the page is updated with the new permissions.  The secure OS
then provides a physical page to the unsecure OS. Since the APU table
update occurs on the trusted interposer, the chiplets are unaware of
the memory allocation request.  \new{Critically, a malicious chiplet
  that somehow gains knowledge of the request still cannot access the
  region, due to the newly set permissions in the APU table, before
  any malicious operation may target the memory region.}

\subsection{Implementation Details}
\label{sec:Implementation}


\subsubsection{NoC Configuration}
Regardless of the interposer's NoC topology, CMCs are emplaced at the
interface between chiplets or MCs and the active interposer.  However,
the width of the physical link does impact the CMC design and its
logic. In our implementation and evaluation, the link width is 128
bits within chiplets and 64/128 bits within the interposer.

\begin{figure}[tb]
  \centering
  \includegraphics[width=0.9\columnwidth]{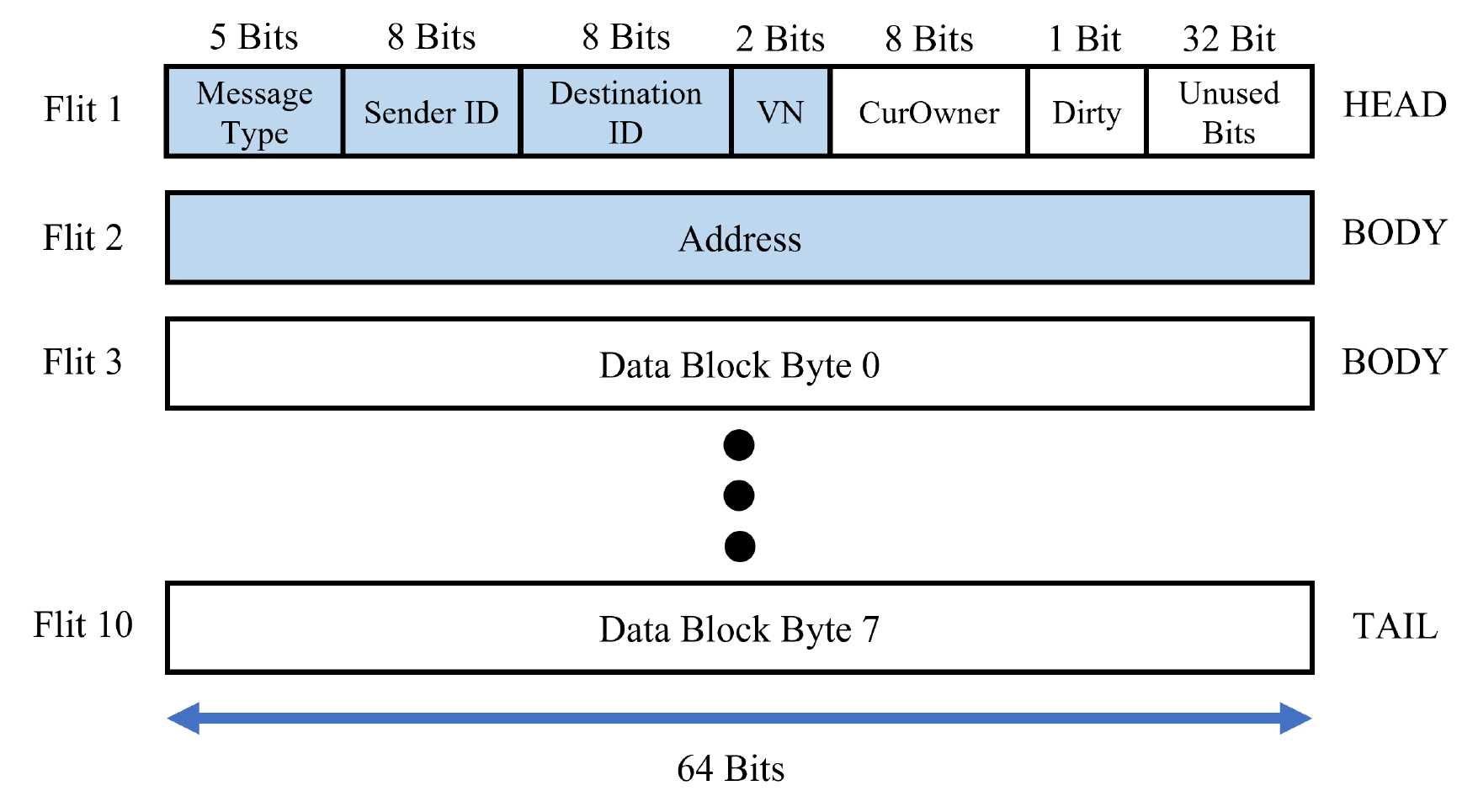}
  \label{fig:Message_Format}
  \caption{Structure of messages. Request messages do not include the `CurOwner'
   or `Dirty' fields. Flits 3-10 are only sent for response messages in response to a request message. 
   Fields highlighted are to be checked by CMCs.}\label{fig:flits}
\smallerspacecaption
\smallerspacecaption
\end{figure}

In MOESI Hammer (Sec.~\ref{sec:cc}), every control message fits within
a single 128-bit flit. When a flit enters the interposer, it is broken
down into two/one flits which are analyzed in the CMC logic over
two/one clock cycles, depending on the 64/128 width of the interposer
link.  In the case of 64-bit links, depicted in Fig.~\ref{fig:flits},
we dedicate the first cycle to extract the control parameters from the
head and the second cycle to extract the address for the cache block
being accessed.  The CMC logic is similar for request and response
messages, as both cases require the first two flits to be analyzed.


\subsubsection{Cache Coherence Protocol}
The system's cache coherence protocol directly impacts the CMC design
and logic as the coherence message fields need to be analyzed by the
CMC.

Note that MOESI Hammer response messages are either a control or data
message; a control response follows the same flit structure as a
request, whereas a data response carries additional flits containing a
total of 64 bytes of data. 

Based on the message type and identified threats (Sec.~\ref{Threat
  Model}), the CMC must analyze certain key parameters; these are
highlighted in Fig.~\ref{fig:flits}.  The parameters are extracted by
the PCM and compared with the permissions set in the APU table. We
note the importance of analyzing both request and response messages,
since an attacker may exploit either message type.

\subsubsection{Protocol Compliance}
\label{sec:malpack}
First, coherence messages are converted into network packets by the
chiplets' NIs. However, these packets are not guaranteed to adhere
to 
the rules of the network and coherence protocols.
For example, a Trojan
may fabricate an invalid
message type, yielding undefined, possibly vulnerable behavior.
Second, messages corresponding to particular virtual networks (VNs) must follow a
specific, limited set of requester/destination IDs and message types.

To address both aspects, the PCM checks the possible field values to
verify the legality of messages.  Since these checks are orthogonal to
memory-region permission checking, they are performed in parallel and
incur no extra delay.

\subsubsection{Design Cost}
\label{sec:overhead}
We design the PCM module with three pipeline stages for lookup, packet
checking, and packet modification.  The third stage is bypassed in
CMC-1 instances as they only monitor packets on ingress to the
interposer.  An APU table requires two bits for identifying each
chiplet's permissions, and there are 64 table entries; 1024 bits are
required for an APU table.  An APU table is in each of the twelve routers (8 
for the chiplets, 4 for the MCs) in the interposer, imposing a total memory 
footprint of 1.5KB.

%% file: evaluation.tex
\section{Evaluation}
\label{Evaluation}

We first discuss our evaluation methodology. Then, we examine the
security coverage our design provides. Finally, we examine the
performance overheads caused by our scheme.
  
\begin{table}[tb]
  \footnotesize
  \centering
  \begin{tabular}{c|c}
    \hline
    \textbf{Component} & \textbf{Variable} \\ \hline \hline
    \multicolumn{2}{c}{\textbf{Chiplet Architecture}}  \\ \hline
    Core & 8 RISC-V cores \\ \hline
    Private L1   I-Cache &32KB \\ \hline
    Private L1   D-Cache &64KB \\ \hline
    L2 Cache & 2MB \\ \hline
    NoC & Eight-port, \\
                       & 128-bit Crossbar \\ \hline
    vc\_per\_vnet &   4, 6, 8, or 10 \\ \hline
    Chiplet Frequency & 1GHz\\ \hline
    \multicolumn{2}{c}{\textbf{System Architecture}} \\ \hline
    Chiplets & 8 \\ \hline
    MCs & 4 \\ \hline
    Main Memory & 4GB  \\ \hline
    Memory Regions & 64, 64MB each \\ \hline
    NoC & 3x4 2D-Mesh, \\
                       & 64 or 128 bit \\ \hline
    vc\_per\_vnet &   4, 6, 8, or 10 \\ \hline
    Interposer Frequency & 250MHz\\ \hline
    \multicolumn{2}{c}{\textbf{Cache Coherence}}  \\ \hline
    Model & AMD MOESI   Hammer\\ \hline
    \multicolumn{2}{c}{\textbf{Simulation Configuration}}  \\ \hline
    Processor Model &TimingSimpleCPU \\ \hline
    Simulation Model &System emulation \\ \hline
  \end{tabular}
  \caption{System Architecture Configuration}
  \label{tab:Parameters of Simulated Architecture}
\smallerspacecaption
\smallerspacecaption
\end{table}

\subsection{Methodology}
We implement and evaluate our proposed system for system emulation using 
gem5~\cite{lowepower2020gem5}. Table~\ref{tab:Parameters of Simulated 
Architecture} depicts the configuration details.
The system is inspired by the Rocket-64 design~\cite{kim19}. Thus, we
simulate an 8-chiplet, 64-core system as described in Sec.~\ref{Arch}. 
The interposer is assumed to be fabricated using an older process node; it operates at 250MHz, a quarter of the chiplets' frequency.

Performance impact is measured as IPC speedup/slowdown for the secure,
CMC-enabled configuration over the unsecure baseline configuration. The CMCs latencies
are discussed in Sec.~\ref{sec:Implementation} and disabled for the unsecured 
baseline. Due to long simulation times induced for this
large system, we evaluate the IPC using a subset of the SPEC 2006
benchmarks.  We perform single-threaded and multi-programmed benchmark
simulations to better understand the impact of the CMCs.

\subsection{Security Analysis}

\subsubsection{Threat Model Coverage}
Our scheme addresses each of the threats discussed in Sec.~\ref{Threat
  Model} as follows:

\textit{Passive reading}: 
  This threat is prevented by rerouting broadcast messages as they 
  enter the CMC-2 located at the interposer/MC boundary.
  Broadcast messages from the  directories are converted into negative 
  acknowledgments back to the requester for chiplets that do not have 
  permissions to the message's memory region.

\textit{Masquerading}:
  Every CMC-1 is programmed with the
  range of ID's expected in each coherence message's
  requester ID field.  For example, in
  Fig.~\ref{fig:System_Architecture}, the CMC-1 in router 72 can
  expect requestor IDs in the range of 0 to 7 and will reject any
  message with an ID outside of this range, as discussed in
  Sec.~\ref{sec:malpack}. In this event, the CMC will throw a
  security check exception and halt execution.

\textit{Modification}: This is detected by comparing a
  message type, such as GETX/GETS, with the access permissions 
  in the APU table. If a message seeks to access memory outside
  of its allowed address space, a security 
  check exception is thrown.

\textit{Diversion}: This threat is detected by checking
  the destination ID and the message type. Only specific message types
  can have other cores as the destination ID. This, along with the
  memory region permissions in the APU table, allows us to detect any
  malicious diversion of messages.  A security
  check exception is thrown if a threat is detected.

Our design generally prevents unauthorized
accesses to memory regions due to privilege escalation or exploitation
using mechanics described above for hardware threats.  Importantly, since coherence messages are generated by hardware, a solely software-driven
attack cannot engage in masquerading, modification, or diversion threats through packet manipulation without malicious hardware intervention.

\subsubsection{Security Testing}
To test the system's ability to counter the
discussed threats, we inject tailored, malicious coherence messages at
the network interface of cores. We verify that, for masquerading,
modification, and diversion, a respective check exception is
thrown and no malicious packets enter the interposer NoC before the
system halts.  For passive reading, recall the \GS\ demonstration
in Sec.~\ref{sec:getx_spy}.

\subsection{Single-Threaded Performance Impact}

\begin{figure}[tb]
  \begin{center}
    \includegraphics[width=.9\columnwidth]{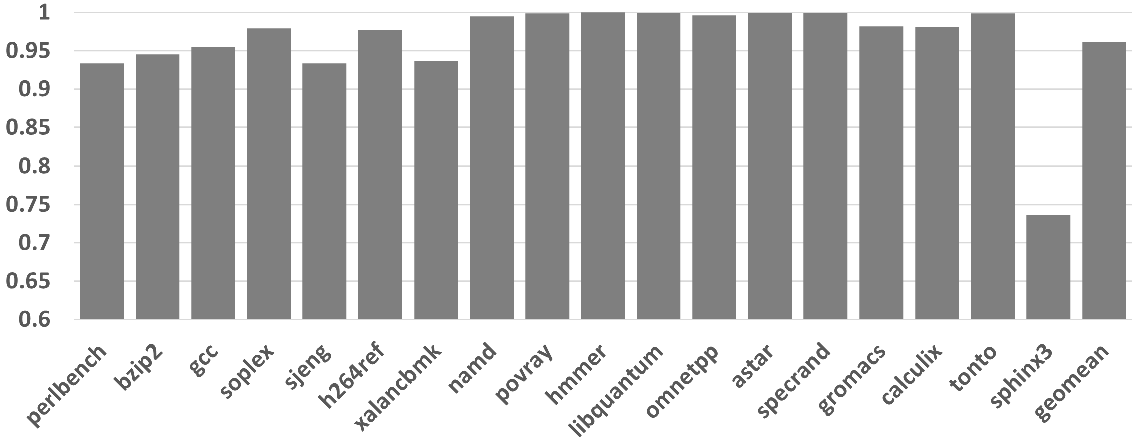}
    \caption{Speedup for the CMC-enabled system, compared to non-secure baseline
      architecture, for vc\_per\_vnet of 4.}
    \label{fig:VC4_speedup}
  \end{center}
\smallerspacecaption
\smallerspacecaption
\end{figure}
  
Figure~\ref{fig:VC4_speedup} shows the speedup/slowdown of the system
with CMCs enabled compared to the baseline configuration.
All workloads experience a speedup less than 1, which is expected, as
the CMCs introduce higher latencies to the network. As the figure
shows, the CMCs impose an average performance loss of $\sim$4\%,
with several benchmarks (\emph{povray}, \emph{hmmer},
\emph{libquantum}) showing little to no impact. \emph{sphinx3},
however, is an outlier, showing a significant $\sim$27\% performance
loss.

\begin{figure}[tb]
  \begin{center}
    \includegraphics[width=.9\columnwidth]{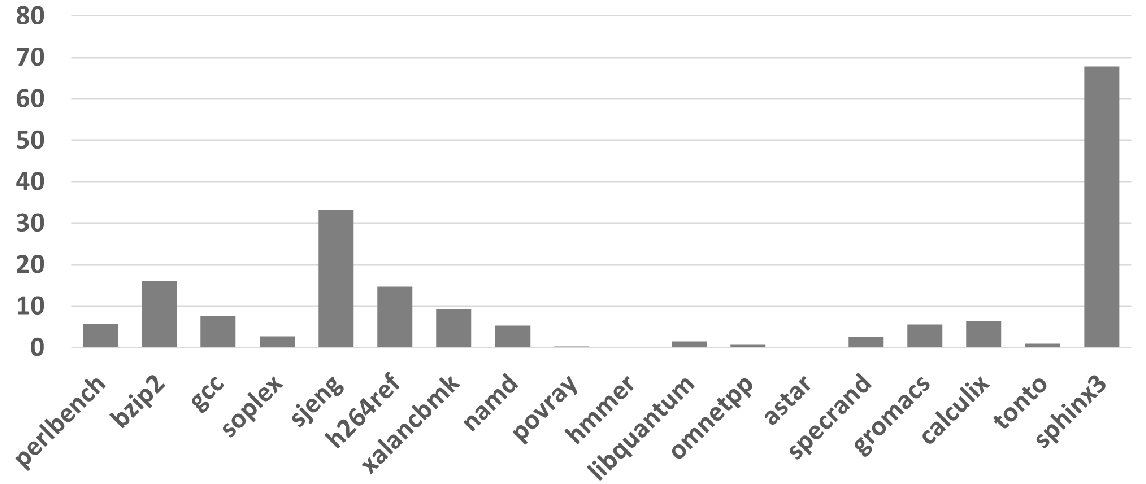}
    \caption{L2 cache misses.}
    \label{fig:L2_miss}
  \end{center}
\smallerspacecaption
\smallerspacecaption
\end{figure}

To analyze further, we examine the L2 miss rates of each
benchmark in Fig.~\ref{fig:L2_miss}.  The figure demonstrates that the variation between each benchmark's result in Fig.~\ref{fig:VC4_speedup} is highly correlated to a benchmark's cache hit rate. 
For instance, \emph{sphinx3} shows a much higher L2 cache miss rate than other benchmarks at $\sim$68\%.
The CMCs must process each packet resulting in increased memory access latencies. Thus, the CMC-enabled system's performance depends on the number
of coherence messages that L2 cache misses inject into the NoC.

\begin{figure}[tb]
  \begin{center}
    \includegraphics[width=.85\columnwidth]{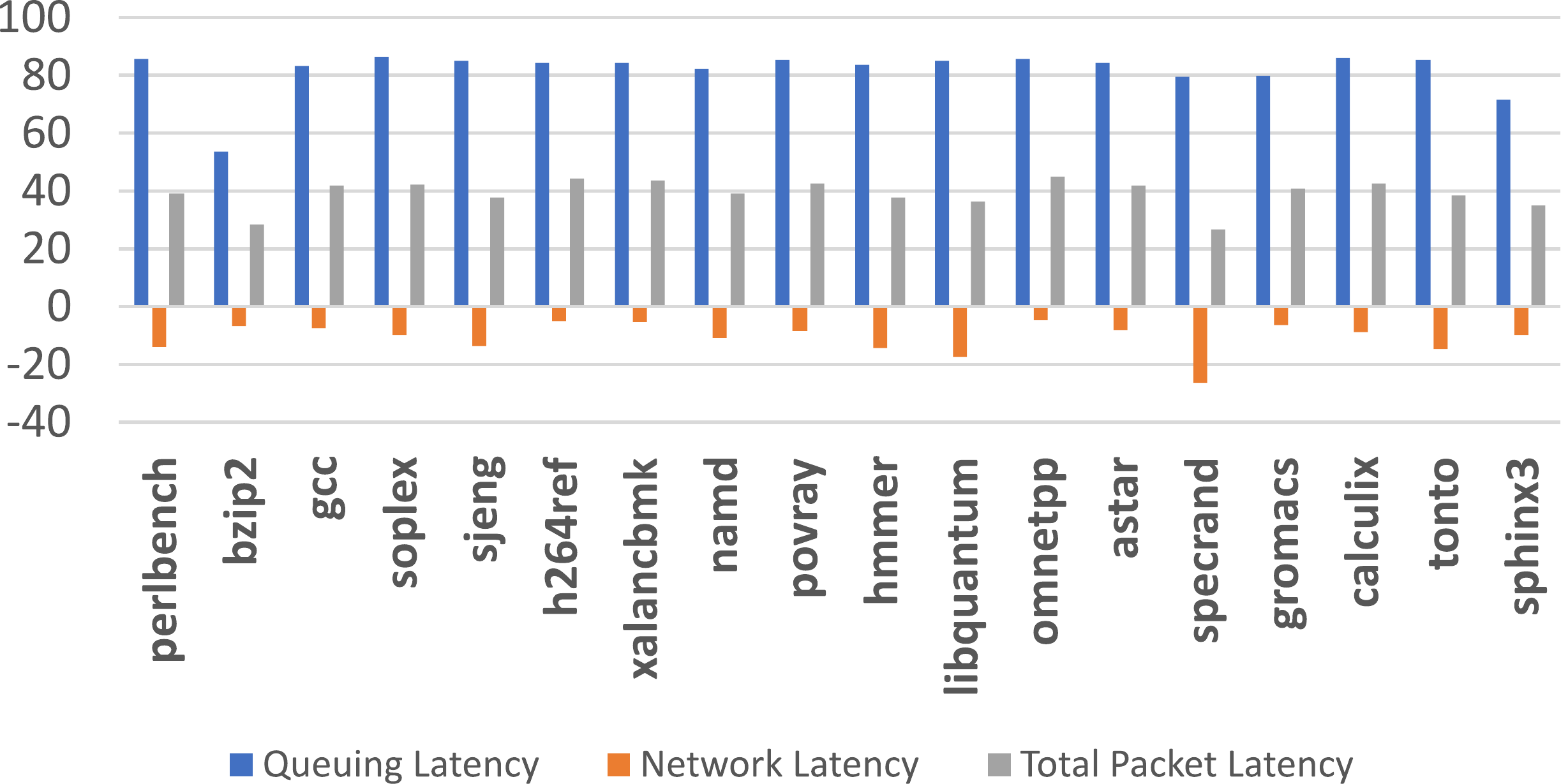}
    \caption{Percent change in packet latency induced by CMCs.}
    \label{fig:Percentage Latency Changes}
  \end{center}
\smallerspacecaption
\smallerspacecaption
\smallerspacecaption
\end{figure}
  
The performance degradation in some benchmarks is analyzed in
Fig.~\ref{fig:Percentage Latency Changes}, showing the percentage change
for pre-injection queuing latency versus in-network latency and the
total latency experienced by packets in the network.
Interestingly, while the queuing latency increases by $\sim$80\%, the
in-network latencies drop by 5--10\%.  The increase in queuing
latency is expected, due to the extra pipeline delays on network
insertion that the CMCs cause.
The decrease in in-network latency is due to
CMC-2 instances rerouting
acknowledgment messages back to only the original requester
(as a negative acknowledgment). Thus, the CMC-2 reduces total network
load by removing one packet in the transaction.

The total packet latency increases by 39\% on average.
Interestingly, although \emph{sphinx3} incurs a higher
performance impact than the other benchmarks, it does not see a
significantly different packet latency. That is, \emph{sphinx3}'s performance
loss is due to a higher L2 miss rate and hence higher packet injection, as
discussed above, not a higher per-packet latency.  Its higher miss
rate exposes \emph{sphinx3} more to the increase of network latency than
other applications, which have lower miss rates.

\begin{figure}[tb]
  \begin{center}
    \includegraphics[width=.9\columnwidth]{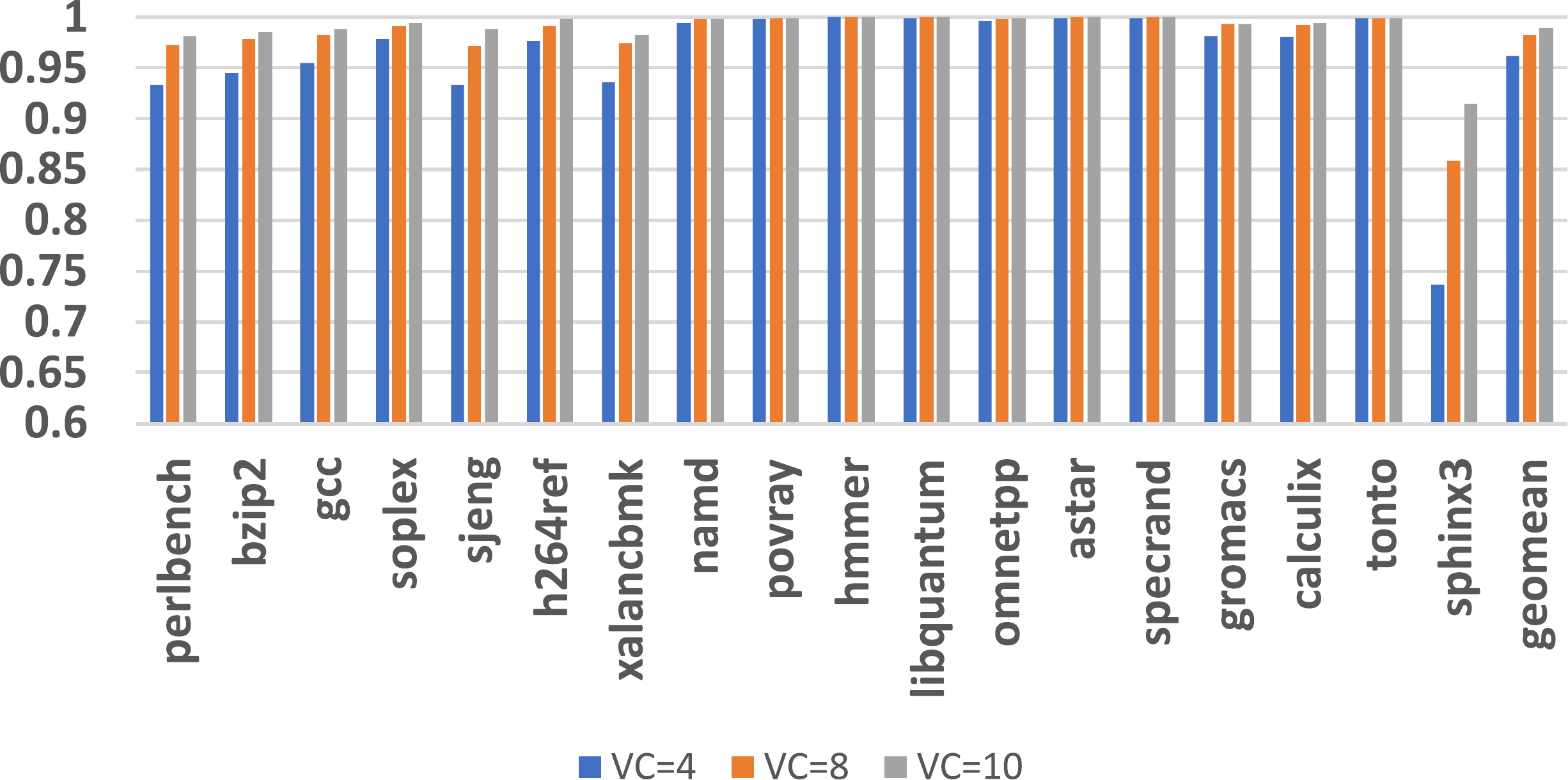}
    \caption{Speedup for different vc\_per\_vnet configurations.}
    \label{fig:All VC Speedup}
  \end{center}
\smallerspacecaption
\smallerspacecaption
\smallerspacecaption
\end{figure}

Figure~\ref{fig:All VC Speedup} depicts the speedup of the benchmarks
with three different  virtual channel configurations (vc\_per\_vnet).
We observe that the geometric mean speedup approaches 0.98 with more virtual channels.
We see a significant improvement
in speedup for \emph{sphinx3} due to the improvement in
queuing latencies at the network interfaces.  These significant gains
imply that increasing VC count is a good way to improve performance
if the application has a high cache miss rate.

\begin{figure}[tb]
  \begin{center}
    \includegraphics[width=.75\columnwidth]{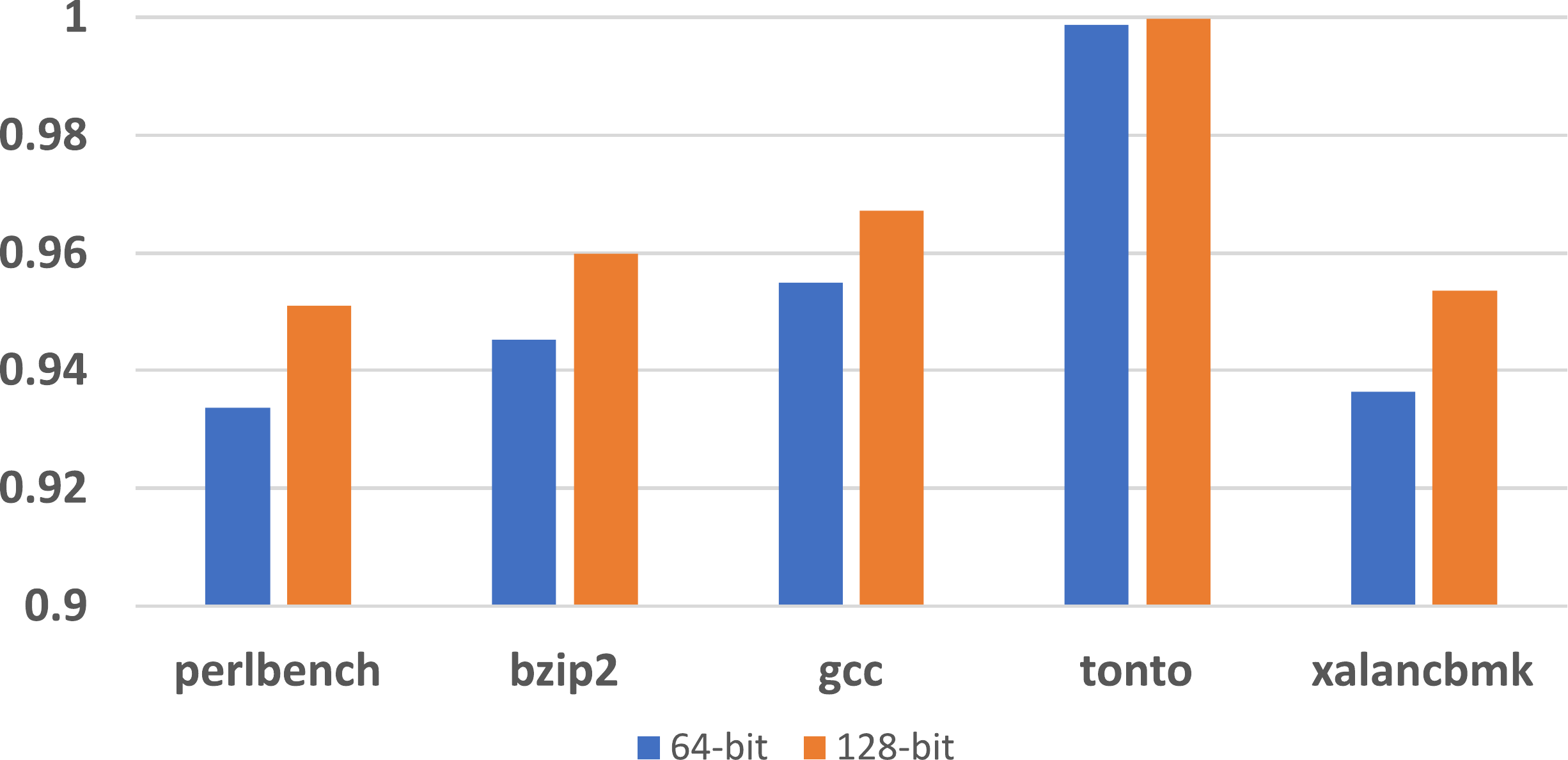}
    \caption{Speedup for 128-bit links within interposer.}
    \label{fig:128_bit_analysis}
  \end{center}
\smallerspacecaption
\end{figure}

In Fig.~\ref{fig:128_bit_analysis}, we analyze the impact of
increasing the interposer link widths to 128 bits versus the baseline
of 64 bits.\footnote{Due to runtime constraints for such large-scale
  simulations running on our shared high-performance computing
  cluster, we focused on a representative subset of benchmark runs for
  that particular experimentation.}  This larger bandwidth provides 
slightly better speedup compared to the baseline.  These modest
gains imply that increasing the bit-width for the physical links in
the interposer is likely not worthwhile, although this depends on the
designer's trade-off for costs/overheads and scalability of the
system.

\subsection{Multi-Programmed Performance Impact}

\begin{figure}[tb]
  \begin{center}
    \includegraphics[width=.8\columnwidth]{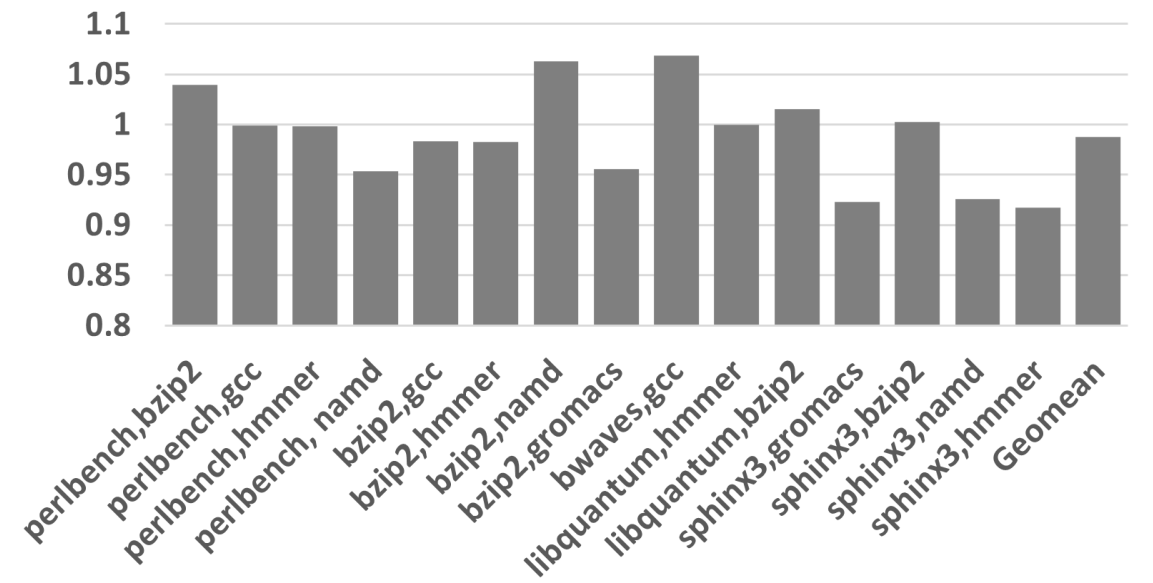}
    \caption{Speedup for multi-programmed workloads.}
    \label{fig:Multi-program}
  \end{center}
\smallerspacecaption
\smallerspacecaption
\smallerspacecaption

\end{figure}

We evaluate the impact of the CMCs for multi-programmed workloads
using random mixes of two benchmarks each, executed in two cores in
separate chiplets.  Here we simulate until all applications complete
at least five billion instructions and we report the weighted speedup
of the combination using a methodology from Kadjo \emph{et
  al.}~\cite{7011422}.

Figure~\ref{fig:Multi-program} shows the speedup for these
multi-programmed workloads.  In general, speedups range between 0.95
and 1.06.  In some cases, namely \emph{bzip2-namd} and
\emph{bwaves-gcc}, the speedup with the CMCs enabled was better than
the baseline.  Further, the mixes which included \emph{sphinx3} showed
reduced performance loss versus the stand-alone \emph{sphinx3}. As before, the improvement is a result of CMC-2 filtering out packets otherwise sent to unauthorized chiplets. This reduces the bandwidth pressure that multiple applications induce on the NoC and appears to reduce the performance overhead as the number of workloads increase.


%% file: conc.tex
\section{Conclusion}
\label{Conclusion}
In this work, we propose the use of an active interposer as root of
trust for modern chiplets-based systems, by implementing hardware
security features directly within the interposer.  More specifically,
we devise a coherence message checker (CMC), which we
propose to include at the boundary between the interposer and the
chiplets/memory controllers. We show how such a scheme addresses
various attacks arising from malicious chiplets,
with relatively low performance impact, $\sim$4\% on average, compared
to a non-secure baseline system.
